\documentclass[%
 aip,
 amsmath,amssymb,
 reprint,%
]{revtex4-2}

\usepackage{dcolumn}
\usepackage{bm}
\usepackage{braket}
\usepackage{color}
\usepackage{natbib}
\usepackage[normalem]{ulem}
\usepackage{ifpdf}
\ifpdf
  \usepackage{graphicx}
\else
  \usepackage[dvipdfmx]{graphicx}
\fi

\begin{document}


\title{
Trapping an Atomic Ion using Time-Division Multiplexed Digital-to-Analog Converters
}

\author{Ryutaro~Ohira}
\email{ohira@quel-inc.com}
\affiliation{
QuEL, Inc., Hachioji ON Building 5F, 4-7-14 Myojincho, Hachioji, Tokyo, Japan
}

\author{Masanari~Miyamoto}
\affiliation{
Graduate School of Engineering Science, The University of Osaka, 1-3 Machikaneyama, Toyonaka, Osaka, Japan
}

\author{Shinichi~Morisaka}
\affiliation{
QuEL, Inc., Hachioji ON Building 5F, 4-7-14 Myojincho, Hachioji, Tokyo, Japan
}
\affiliation{
Center for Quantum Information and Quantum Biology, The University of Osaka, 1-2 Machikaneyama, Toyonaka, Osaka, Japan
}

\author{Ippei~Nakamura}
\affiliation{
Komaba Institute for Science (KIS), The University of Tokyo, 3-8-1 Komaba, Meguro, 153-8902, Tokyo, Japan
}

\author{Atsushi~Noguchi}
\affiliation{
Komaba Institute for Science (KIS), The University of Tokyo, 3-8-1 Komaba, Meguro, 153-8902, Tokyo, Japan
}
\affiliation{
RIKEN Center for Quantum Computing (RQC), 2-1 Hirosawa, Wako, 351-0198, Saitama, Japan
}
\affiliation{
Inamori Research Institute for Science (InaRIS), 620 Suiginya, Kyoto, 600-8411, Kyoto, Japan
}

\author{Utako~Tanaka}
\affiliation{
Graduate School of Engineering Science, The University of Osaka, 1-3 Machikaneyama, Toyonaka, Osaka, Japan
}
\affiliation{
Center for Quantum Information and Quantum Biology, The University of Osaka, 1-2 Machikaneyama, Toyonaka, Osaka, Japan
}
\affiliation{
National Institute of Information and Communications Technology, 588-2, Iwaoka, Nishi-ku, Kobe, Hyogo, Japan
}

\author{Takefumi~Miyoshi}
\affiliation{
QuEL, Inc., Hachioji ON Building 5F, 4-7-14 Myojincho, Hachioji, Tokyo, Japan
}
\affiliation{
Center for Quantum Information and Quantum Biology, The University of Osaka, 1-2 Machikaneyama, Toyonaka, Osaka, Japan
}
\affiliation{
e-trees. Japan, Inc., Daiwaunyu Building 2F, 2-9-2 Owadamachi, Hachioji, Tokyo, Japan
}

\date{\today}

\begin{abstract}
Independent control of numerous electrodes in quantum charge-coupled device architectures presents a significant challenge for wiring and hardware scalability. 
To address this issue, we demonstrate a voltage control method based on time-division multiplexing (TDM).
This approach utilizes a single high-update-rate digital-to-analog converter (DAC) to sequentially generate control signals for multiple electrodes, thereby reducing both the number of required DACs and associated wiring.
We experimentally validate this concept by developing a 10-channel system that operates with only two DACs.
The developed TDM-based voltage control system is applied to a surface-electrode trap, where we successfully trap a single $^{40}\mathrm{Ca}^+$ ion and demonstrate a simple ion transport primitive.
This approach offers a resource-efficient and scalable solution for advanced quantum computing systems based on trapped ions.
\end{abstract}

\maketitle

The quantum charge-coupled device (QCCD) architecture is a leading candidate for realizing large-scale trapped-ion quantum computers~\cite{wineland1998experimental, kielpinski2002architecture, lekitsch2017blueprint, kaushal2020shuttling, pino2021demonstration, moses2023race, mordini2025multizone}. 
It relies on ion traps with numerous segmented electrodes, which enable ion shuttling---the transport of qubits between different processing zones via time-dependent voltage control~\cite{kaushal2020shuttling}.

As QCCD systems scale, the independent control of a vast number of electrodes becomes a formidable challenge, directly translating into a proportional increase in classical control electronics and wiring complexity~\cite{malinowski2023wire}. 
For instance, operating a system with only a few dozen qubits already requires hundreds of DC control signals~\cite{moses2023race}. 
Each trapped-ion qubit typically necessitates multiple electrodes for basic operations such as trapping, transport, splitting, merging, and rotation, leading to a rapid increase in the number of independent control lines as system size grows. 
This trend imposes practical limitations related to vacuum chamber feedthroughs, physical wire routing, and the volume of classical electronics required for signal generation.

Several strategies have been investigated to address this bottleneck. 
One approach reduces the number of required DC signals through voltage broadcasting, where a single signal is used to drive multiple electrodes~\cite{moses2023race, delaney2024scalable}. 
In particular, Delaney et al.~\cite{delaney2024scalable} combined voltage broadcasting with per-trap-site voltage switching, implementing a control primitive, that lowers the total number of control lines.
Another strategy involves relocating parts of the control electronics inside the vacuum chamber, thereby minimizing external connections and simplifying the wiring infrastructure~\cite{guise2014vacuum, stuart2019chip, malinowski2023wire}.

Recently, we proposed an alternative method based on time-division multiplexing (TDM), employing a high-update-rate digital-to-analog converter (DAC) for scalable electrode control~\cite{ohira2025multiplexed}.
While conventional QCCD systems typically allocate one DAC---operating at a few mega-updates per second (MUPS)---to each electrode~\cite{blakestad2010transport, blakestad2011near, akhtar2023high, kaushal2020shuttling, mordini2025multizone}, our method employs fewer high-speed DACs that sequentially generate voltage signals for multiple electrodes.
These signals are routed via a demultiplexer to the corresponding electrodes, thereby reducing both the number of DACs and the overall wiring complexity.

In our previous work~\cite{ohira2025multiplexed}, we proposed a TDM-based control scheme and demonstrated only a preliminary proof-of-concept with five multiplexed channels, without application to an actual ion trap.
In this work, we present an experimental proof of concept for a TDM-based approach to scalable electrode control.
We develop a system capable of controlling ten independent trap electrodes using only two DACs, and validate its functionality by trapping a single $^{40}\mathrm{Ca}^+$ ion in a surface-electrode trap.
We also demonstrate a simple ion transport primitive.
This result demonstrates the potential of TDM-based scalable electrode control as a practical solution for reducing classical hardware overhead in large-scale QCCD systems.

\begin{figure}[t]
    \centering
    \includegraphics[width=8.5cm]{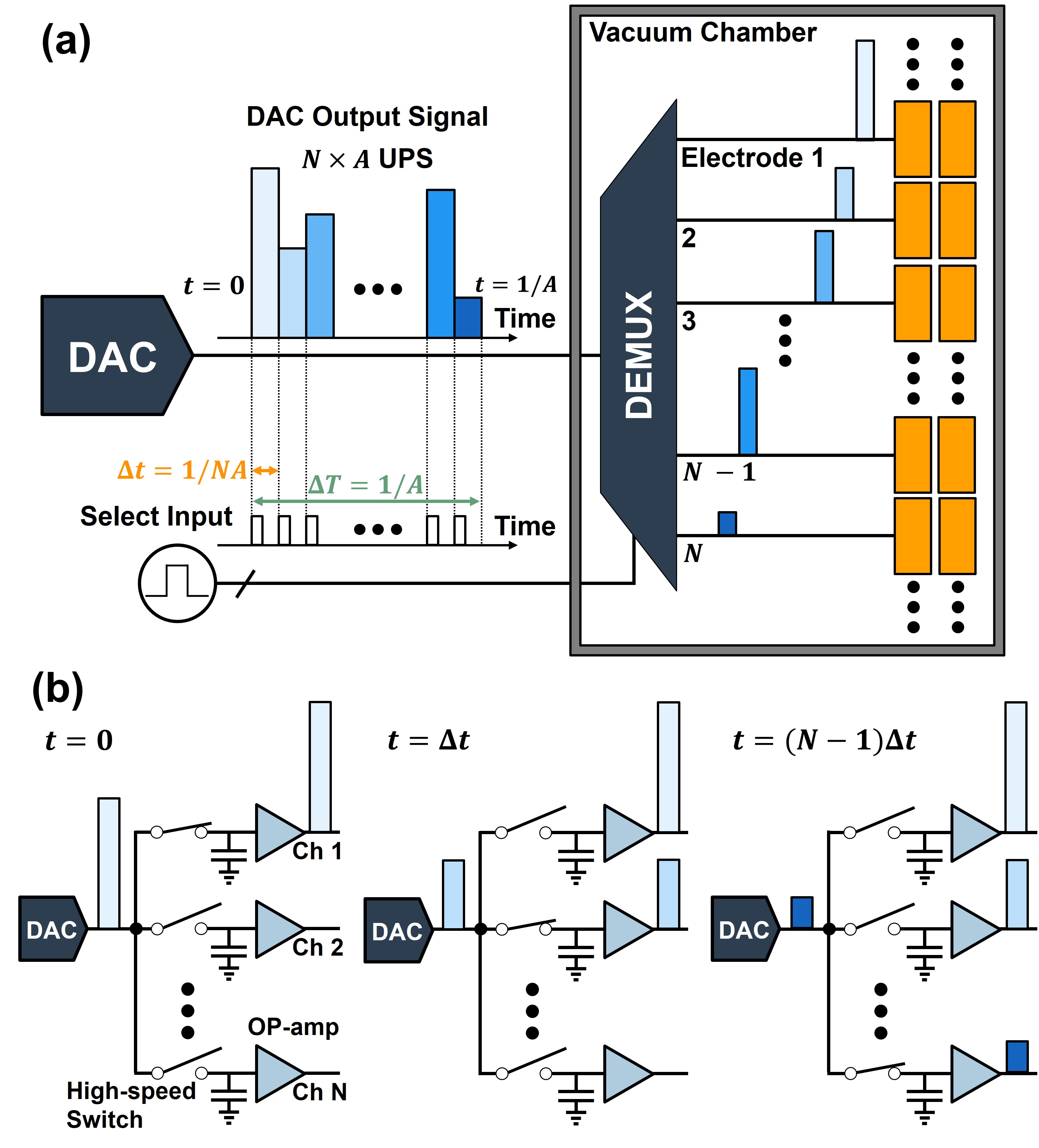}
    \caption{
    (a) Conceptual diagram of the proposed control architecture. A TDM signal is generated by a DAC operating at an update rate of \( N \times A \) updates per second (UPS), where \( A \) is the effective update rate per output channel and \( N \) is the number of channels. This multiplexed signal is transmitted to an in-vacuum demultiplexer (DEMUX), whose operation is governed by a logic signal labeled ``Select Input''.
    (b) Implementation of the demultiplexer. The design employs high-speed switches and voltage-storage capacitors. The switches, controlled by the Select Input logic signal, sequentially route the TDM signal to the capacitors. For example, Channel 1's switch is enabled at $t=0$, Channel 2's at $t=\Delta t$, and Channel $N$'s at $t=(N-1)\Delta t$, where $\Delta t = 1/NA$. Each capacitor retains its voltage for $\Delta T = N\Delta t = 1/A$ seconds between updates, reconstructing the desired waveform for each electrode. If needed, the signal is amplified to the target voltage using an operational amplifier.
    }
    \label{Fig:concept}
\end{figure}

The TDM-based electrode control scheme, originally proposed in Ref.~\cite{ohira2025multiplexed}, forms the basis of the experimental implementation presented here.
This method synthesizes TDM signals corresponding to the desired voltage waveforms and distributes them to individual output channels via a demultiplexer, thereby reconstructing the appropriate control voltages at each electrode.
As shown in Fig.~\ref{Fig:concept}(a), a single DAC generates a TDM signal at an update rate of \( N \times A \) updates per second (UPS), where \( A \) is the effective update rate per output channel and \( N \) is the number of channels.  
The multiplexed signal is then delivered to a demultiplexer housed within the vacuum chamber, significantly reducing the number of physical feedthroughs required between room-temperature electronics and the vacuum environment.
The demultiplexer, labeled ``DEMUX'' in Fig.~\ref{Fig:concept}(a), is controlled by a logic signal, labeled ``Select Input'' in Fig.~\ref{Fig:concept}(a), which is synchronized with the DAC's voltage update rate.

As detailed in Ref.~\cite{ohira2025multiplexed} and illustrated in Fig.~\ref{Fig:concept}(b), demultiplexing is achieved using high-speed switches in conjunction with voltage-storage capacitors.
The Select Input logic signal dictates the on/off state of each switch, allowing the system to cycle through $N$ channels at a rate of $N\times A$~Hz. 
Each capacitor is charged to its target voltage during a narrow time window of $\Delta t = 1/NA$ when its corresponding switch is enabled. 
For instance, as shown in Fig.~\ref{Fig:concept}(b), Channel 1's switch activates at $t=0$, Channel 2's at $t=\Delta t$, so forth, with Channel $N$ activating at $t=(N-1)\Delta t$.
The cycle repeats every $\Delta T=N\Delta t=1/A$ seconds, ensuring periodic updates of all channel voltages.

During this interval of $\Delta T$, the capacitor must maintain its stored voltage with minimal discharge.
Any significant voltage drift may adversely affect the motional states of the ions and reduce the fidelity of quantum gate operations.
Note that this technique draws inspiration from the architecture introduced in Ref.~\cite{malinowski2023wire}, which also suggested that incorporating a digital decoder could further reduce the number of required digital control lines from $N$ to $log_2 N$.
Finally, the reconstructed signal is amplified to the desired voltage level using an operational amplifier (OP-amp), if necessary.

The maximum achievable multiplexing factor---defined as the number of output channels controlled by a single DAC---is constrained by several factors.  
These include the DAC's settling time, the switching speed of the switches, and the charge and discharge time constants of the capacitors.  
As estimated in Ref.~\cite{ohira2025multiplexed}, assuming an acceptable update rate of 0.5~MUPS per channel, a multiplexing factor of approximately 100 is feasible under realistic operating conditions.  
Further technical details and design considerations are discussed in Ref.~\cite{ohira2025multiplexed}.
Provided that sufficient FPGA pins are available, multiple high-update-rate DAC units can be incorporated to expand the number of controllable electrodes.
Additionally, the system architecture is inherently scalable: multiple FPGA modules can be operated in synchronization to increase the overall channel count~\cite{ohira2025multiplexed}.
A practical method for such synchronization involves Gigabit Ethernet (GbE)-based coordination between FPGA units, as explored in Ref.~\cite{miyoshi2025toward}.

\begin{figure*}[t]
    \centering
    \includegraphics[width=17.0cm]{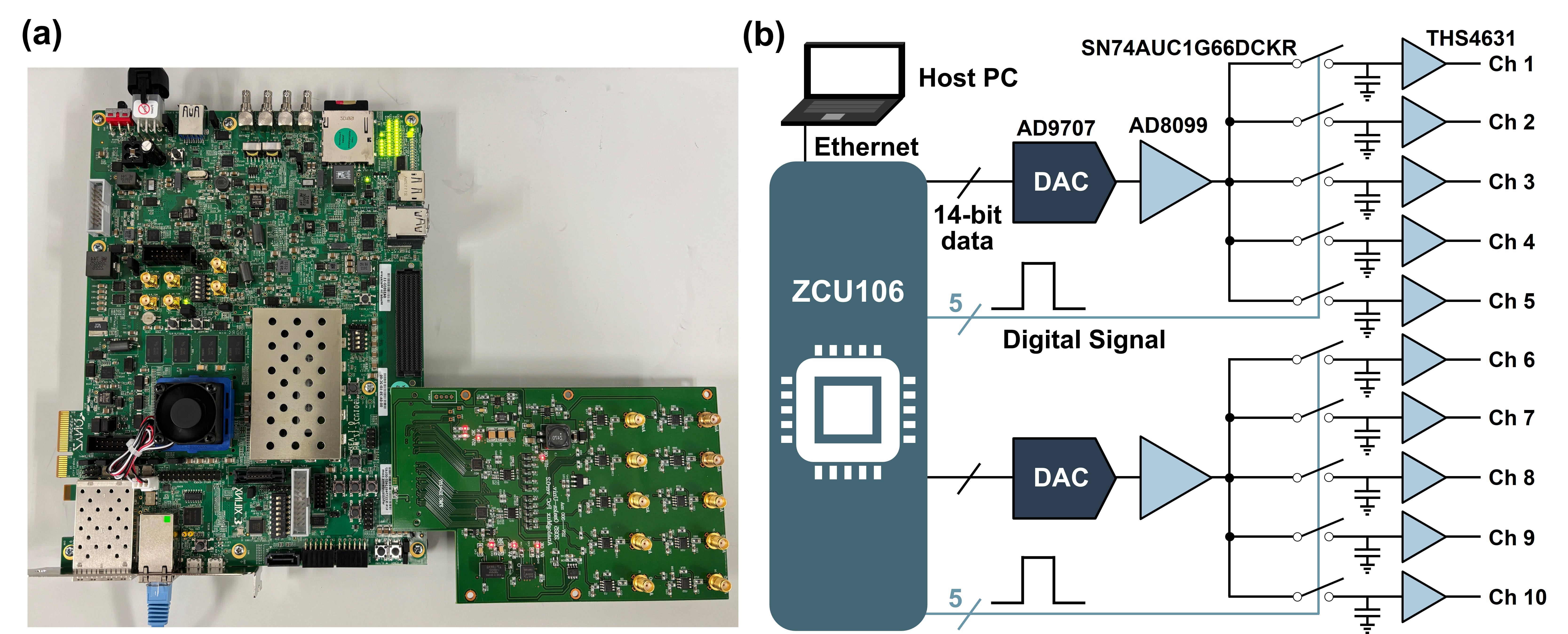}
    \caption{
    (a) Photograph of the developed TDM-based voltage control system.
    (b) Block diagram of the system architecture. The system employs two 14-bit current-output DACs (AD9707), each time-multiplexing five analog control signals. These DACs are driven by an FPGA module (ZCU106), which generates the required TDM signals. Control operations are initiated from a host PC via Ethernet by executing Python scripts. The DAC output currents are converted to voltage signals using current-to-voltage converters based on OP-amps (AD8099). The resulting voltage signals are sequentially routed through high-speed switches (SN74AUC1G66DCKR) to charge dedicated capacitors. These capacitors retain the control voltages between TDM updates. Finally, the stored voltages are amplified by a second-stage OP-amp (THS4631).
    }
    \label{Fig:dev_system}
\end{figure*}
\begin{figure}[t]
    \centering
    \includegraphics[width=8.5cm]{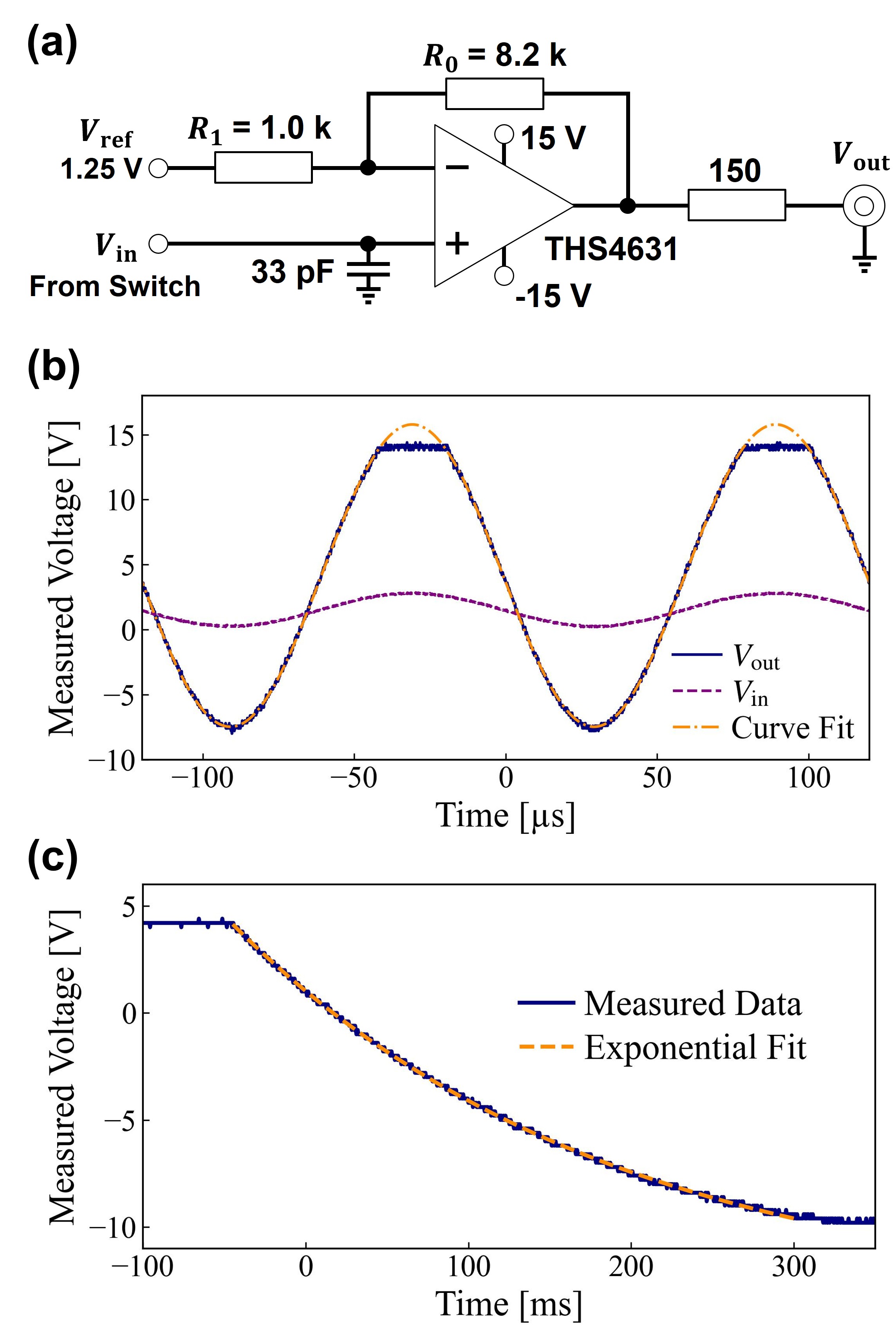}
    \caption{
    (a) Circuit diagram of the second-stage voltage amplifier, using an OP-amp (THS4631) and a fixed reference voltage of 1.25~V.
    (b) Measured waveforms of a full-range sine signal on channel 1. The dashed curve shows $V_{\text{in}}$ immediately after the analog switch, while the solid curve shows $V_{\text{out}}$ after final-stage amplification. Clipping is observed near 14.2~V. A sinusoidal fit to the unclipped region (dash-dot curve) estimates the peak voltage at 15.8~V.
    (c) Measured voltage hold behavior at the output of channel 1. An exponential fit to the decay between -45~ms and 300~ms yields a time constant of 233(1)~ms. The time axes in (b) and (c) represent raw oscilloscope time values.
    }
    \label{fig:characterization}
\end{figure}

Figure~\ref{Fig:dev_system} shows a photograph of the developed TDM-based control system and a schematic of its detailed architecture. 
The system is designed to control ten independent trap electrodes using two 14-bit DACs (AD9707, Analog Devices, Inc.), with each DAC time-multiplexing five analog control signals.
Thus, the system comprises two identical units, each consisting of a DAC and a demultiplexer, as illustrated in Fig.~\ref{Fig:dev_system}(a).
These DACs are driven by a field-programmable gate array (FPGA) module, ZCU106~\footnote{Zynq UltraScale+ MPSoC ZCU106 Evaluation Kit, Advanced Micro Devices, Inc.}, which generates the required TDM signals.
The Zynq device on the ZCU106 includes of a processing system (PS) and programmable logic (PL).
In this configuration, the PL generates 14 parallel digital signals to control the DACs.
Waveform data are prepared in the Linux-based PYNQ environment running on the PS and written into memory allocated in the PL via a C++ driver program.
The control logic implemented in the FPGA then transfers the data to the DACs.
All operations can be initiated from a host PC over an Ethernet connection by executing Python scripts within a Jupyter Notebook hosted on the PYNQ platform.

The AD9707 DAC is interfaced with the FPGA via a 14-bit parallel bus and synchronized using a dedicated clock signal generated by the FPGA. 
As a current-output DAC, the AD9707 feeds its output into a current-to-voltage converter implemented using an OP-amp (AD8099, Analog Devices, Inc.).
This stage converts the current signal into a voltage in the 0--2.5~V range, hereafter denoted as $V_\text{in}$.

In this study, the DACs operate at a voltage update rate of 30~MUPS, resulting in an effective voltage update rate of 6~MUPS for each of the five output channels.
As previously described, this corresponds to a time slot of $\Delta t \approx 33.3$~ns allocated to each channel within a TDM cycle, during which the associated capacitor must be charged.
The interval between successive updates for any given output channel is $\Delta T = 5\Delta t \approx 167$~ns, representing the full TDM cycle across the five channels. 

The output signals from the AD8099 are routed through a demultiplexer, which distributes the generated waveform to individual electrode channels. 
Each demultiplexing unit consists of a high-speed analog switch (SN74AUC1G66DCKR, Texas Instruments Inc.) and a 33~pF capacitor.  
These high-speed switches operate synchronously with the DAC update rate, enabling the time-division distribution of the multiplexed signals.  
Each selected switch connects the DAC output to its corresponding capacitor, charging it to the appropriate voltage. 
Given the on-resistance of the switch is 9~$\Omega$~\cite{SN74AUC1G66-datasheet}, the resulting RC time constant is 297~ps, allowing the capacitor to reach over 99~\% of its target value within approximately 1.5~ns.
This charge-hold mechanism completes the demultiplexing operation by storing the control voltage for each channel.  

The stored voltage is subsequently amplified by a second-stage OP-amp (THS4631, Texas Instruments Inc.) to achieve the final output range.
The detailed circuit diagram is shown in Fig.~\ref{fig:characterization}(a).
The non-inverting input of the OP-amp directly receives the signal $V_\text{in}$, while the inverting input is connected to the amplifier output through a feedback resistor $R_0=8.2~\rm{k\Omega}$ and to a fixed reference voltage $V_\text{ref}=1.25$~V via a resistor $R_1=1.0~\rm{k\Omega}$
This configuration yields a voltage gain of $1+R_0/R_1=9.2$.
Because the inverting input is biased with $V_\text{ref}$, an offset is introduced at the output, resulting in the following expression for the output voltage: 
\begin{equation}\label{eq:vol_cal}
V_{\text{out}}=9.2 \cdot V_{\text{in}} - 8.2 \cdot V_\text{ref}.  
\end{equation}
For example, when $V_{\text{in}} = 0$~V, the output is $V_{\text{out}} \approx -10$~V, and when $V_{\text{in}} = 2.5$~V, it becomes $V_{\text{out}} \approx 13$~V.  

We summarize here the key updates introduced in the present system compared to the earlier implementation reported in Ref.~\cite{ohira2025multiplexed}.
First, the current system supports ten electrode channels by employing two DACs, each managing five control signals.
In contrast, the previous system supported only five channels.
Second, the final-stage OP-amp was selected for its high input impedance, significantly extending the voltage hold time at the output.
This improvement is characterized later in this section.

We now present the electrical characterization of the developed TDM-based control system.
Unless otherwise specified, all measurements reported below were performed on channel 1.
Figure~\ref{fig:characterization}(b) shows the output waveform of a full-range sine signal generated by sweeping the 14-bit DAC code across its entire range.
The waveform is recorded at two points: immediately after the analog switch ($V_{\text{in}}$) and after the final-stage OP-amp ($V_{\text{out}}$).
The measured $V_{\text{out}}$ [solid curve in Fig.~\ref{fig:characterization}(b)] ranges approximately from -7.5~V to 14.2~V, with noticeable clipping observed near the upper bound.
The dash-dot curve represents sinusoidal fit to the unclipped portion of $V_{\text{out}}$, estimating the peak output voltage to reach 15.8~V.
If this fitted peak is taken as representative, and the clipping at 14.2~V is disregarded, the effective output voltage range is -7.5~V to 15.8~V, which deviates from the expected range based on the design.

A likely cause of this deviation is an offset in the input voltage $V_{\text{in}}$.
The measured $V_{\text{in}}$ [dashed curve in Fig.\ref{fig:characterization}(b)] spans from 0.26~V to 2.79~V, deviating from the ideal range of 0~V to 2.5~V.
Substituting these measured values into Eq.~(\ref{eq:vol_cal}), the corresponding $V_{\text{out}}$ is calculated to range from -7.9(3)~V to 15.4(5)~V.
The uncertainties represent $1\sigma$ values, derived by propagating the $\pm 5\%$ tolerances of $R_0$ and $R_1$ under a uniform distribution~\footnote{
$R_0$ and $R_1$ are general-purpose chip resistors, RK73B1JTTD822J and RK73B1JTTD102J, respectively, manufactured by KOA Speer Electronics, Inc. 
According to their datasheet, both resistors have a tolerance of $\pm 5\%$.
}.
The results are consistent with the presence of an input offset, suggesting it is the primary contributing factor.
We find that the offset originates from the regulator that supplies the reference voltage for the first-stage OP-amp (AD8099), which is designed to produce 0--2.5~V under nominal conditions.
The non-inverting input of this OP-amp is tied to a reference voltage obtained by resistive division of the regulator output, nominally 0.80~V.
However, the regulator does not operate properly and outputs 5.43~V; the reference voltage rises to 0.89~V.
Consequently, the output range shifts from the intended 0--2.5~V to approximately 0.28--2.78~V, consistent with the measured value of 0.26--2.79~V.

Figure.~\ref{fig:characterization}(c) presents the measured voltage hold behavior, yielding a time constant of 233(1)~ms.
This value is obtained by fitting an exponential decay function to the data recorded between -45 ms and 300~ms in Fig.~\ref{fig:characterization}(c).
Given this time constant, the corresponding voltage drop over the recharge interval ($\approx$167~ns) is on the order of $10^{-7}$ relative to the initial voltage, which is negligible for practical purposes.

In addition, the slew rate is evaluated to be 193~V/$\mathrm{\mu}$s during transitions across the full DAC code range (14.2~V to -7.5~V). 
This value is derived from the time required for the output to traverse from 10\% to 90\% of the full voltage swing.

\begin{figure}[t]
    \centering
    \includegraphics[width=8.5cm]{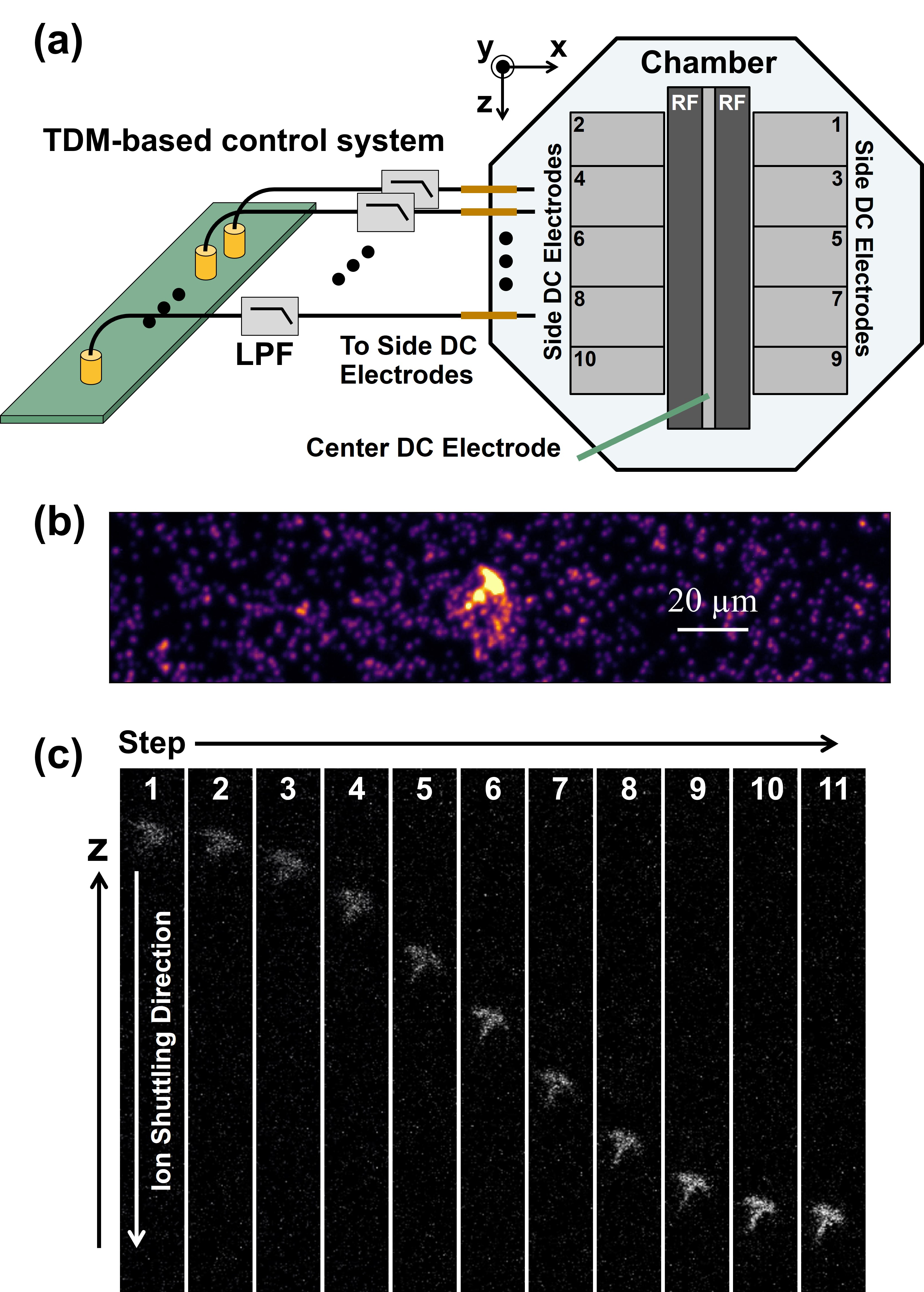}
    \caption{
    (a) Schematic of the experimental setup (not to scale), showing the surface-electrode trap with two RF electrodes and eleven DC electrodes. The developed TDM-based voltage control system drives the ten side DC electrodes. Control voltages pass through low-pass filters (LPFs) before entering the vacuum chamber via vacuum feedthroughs and reaching the electrodes.
    (b) Fluorescence image of a single $^{40}\mathrm{Ca}^+$ ion trapped in the surface electrode trap. Each pixel corresponds to approximately 0.4~$\mu$m.
    (c) Shuttling of a single $^{40}\mathrm{Ca}^+$ ion. The white arrow indicates the ion shuttling direction along the $-z$ axis. The ion is transported by a time-dependent voltage generated by the TDM-based control system and applied to ten side DC electrodes.
    The applied waveform consists of eleven discrete voltage steps with an update interval of 500~ms.
    }
    \label{fig:result}
\end{figure}
\begin{table}[t]
\caption{\label{tab:dc_voltages}%
Assignment of output channels to side DC electrodes and the corresponding applied voltages.
}
\begin{ruledtabular}
\begin{tabular}{ccc}
\textrm{Output Channel} & \textrm{Side DC Electrode No.} & \textrm{Voltage [V]} \\
\hline
Ch 1 & 1 & 0.0 \\
Ch 2 & 2 & 0.0 \\
Ch 3 & 3 & 10.4 \\
Ch 4 & 4 & 10.4 \\
Ch 5 & 5 & 0.8 \\
Ch 6 & 6 & 0.0 \\
Ch 7 & 7 & 10.4 \\
Ch 8 & 8 & 10.4 \\
Ch 9 & 9 & 0.0 \\
Ch 10 & 10 & 0.0 \\
\end{tabular}
\end{ruledtabular}
\end{table}

We now present the experimental application of the developed TDM-based voltage control system to an ion trap.
Figure~\ref{fig:result}(a) illustrates the experimental setup used in this study.
The experiments are conducted using with a surface-electrode trap, the geometry of which is also depicted in Fig.~\ref{fig:result}(a).
Detailed specifications of the trap are available in Ref.~\cite{miyamoto2025isotope}. 
The trap features two RF electrodes and eleven DC electrodes. 
Among the DC electrodes, one is located at the center of the trap (``center DC electrode''), while the remaining ten are symmetrically arranged around the periphery (``side DC electrode''). 

The ten side DC electrodes are driven by the ten output channels of the TDM-based control system.
The center DC electrode is independently controlled by a separate voltage source. 
For radial confinement, an RF signal with a peak-to-peak amplitude of 189~V at 24.3~MHz is applied to the RF electrodes. 

The trap assembly is housed in an ultra-high vacuum chamber maintained at a pressure of approximately 9.8$\times10^{-8}$~Pa. 
The voltages generated by the TDM-based control system are passed through low-pass filters (LPFs) with a cutoff frequency of approximately 2~kHz~\cite{oshio2025development}, then routed through vacuum feedthroughs to the trap electrodes, as illustrated in Fig.~\ref{fig:result}(a).
As previously discussed, the switching array should ideally be located inside the vacuum chamber to minimize the number of feedthroughs. 
However, the primary objective of this study is to demonstrate ion trapping using the TDM-based control scheme. 
Thus, all electronics remain outside the vacuum chamber for the current implementation.

Ion loading begins with a beam of neutral calcium atoms introduced through a hole at the center of the trap~\cite{miyamoto2025isotope}. 
These atoms are photoionized using lasers at 423 nm and 375 nm.
The resulting $^{40}\mathrm{Ca}^+$ ions are Doppler-cooled using lasers at 397~nm and 866~nm.
Ion fluorescence is detected using a CMOS camera, with signal enhancement provided by an image intensifier. 
Further details of the optical setup for ion loading, cooling, and detection are provided in Ref.~\cite{miyamoto2025isotope}.

Figure~\ref{fig:result}(b) presents a fluorescence image of a single $^{40}\mathrm{Ca}^+$ ion trapped using the developed TDM-based voltage control system.
The voltages applied to the side DC electrodes during this experiment are listed in Table~\ref{tab:dc_voltages}.
The center DC electrode is biased at 5.5~V using a separate voltage source.
The trap frequency along the $z$ axis is determined to be approximately 0.65~MHz.
This value is obtained by applying a sinusoidal voltage to the center electrode and observing the resulting fluorescence, which reflects the ion's motional extent.
These results confirm that the TDM-based control scheme is capable of generating suitable trapping potentials for ion confinement.
Notably, the ion is successfully trapped using only two DACs to control ten electrodes, in contrast to conventional systems that typically require one DAC per electrode.

Furthermore, shuttling of a single trapped ion is performed by applying a time-dependent voltage generated by the TDM-based control system.
The experimental conditions differ from those used for static trapping of a single ion.
In this experiment, the RF confinement voltage is set to 240~V, and the vacuum pressure is 4.1$\times10^{-7}$~Pa.
A qCMOS camera (C15550-20UP, Hamamatsu Photonics) is used to record the ion images.
The result of the shuttling experiment is shown in Fig.~\ref{fig:result}(c).
The image shows the ion being transported along the $-z$-axis direction when the time-dependent voltage is applied to the side DC electrodes.
The applied waveform consists of eleven discrete voltage steps, each updated every 500~ms.
The waveform for the ion transport is obtained using the numerical method described in Ref.~\cite{oshio2025development}.
Note that the voltage update rate is relatively slow; the shuttling speed is not optimized in this demonstration.
Here, we simply show that the shuttling primitive can be realized using the TDM-based voltage control scheme.
The center DC electrode is held at a fixed bias of 0.24~V.
During the shuttling operation, both the 397~nm and 866~nm laser beams are kept continuously on along the $z$-axis.

\begin{figure}[t]
    \centering
    \includegraphics[width=8.5cm]{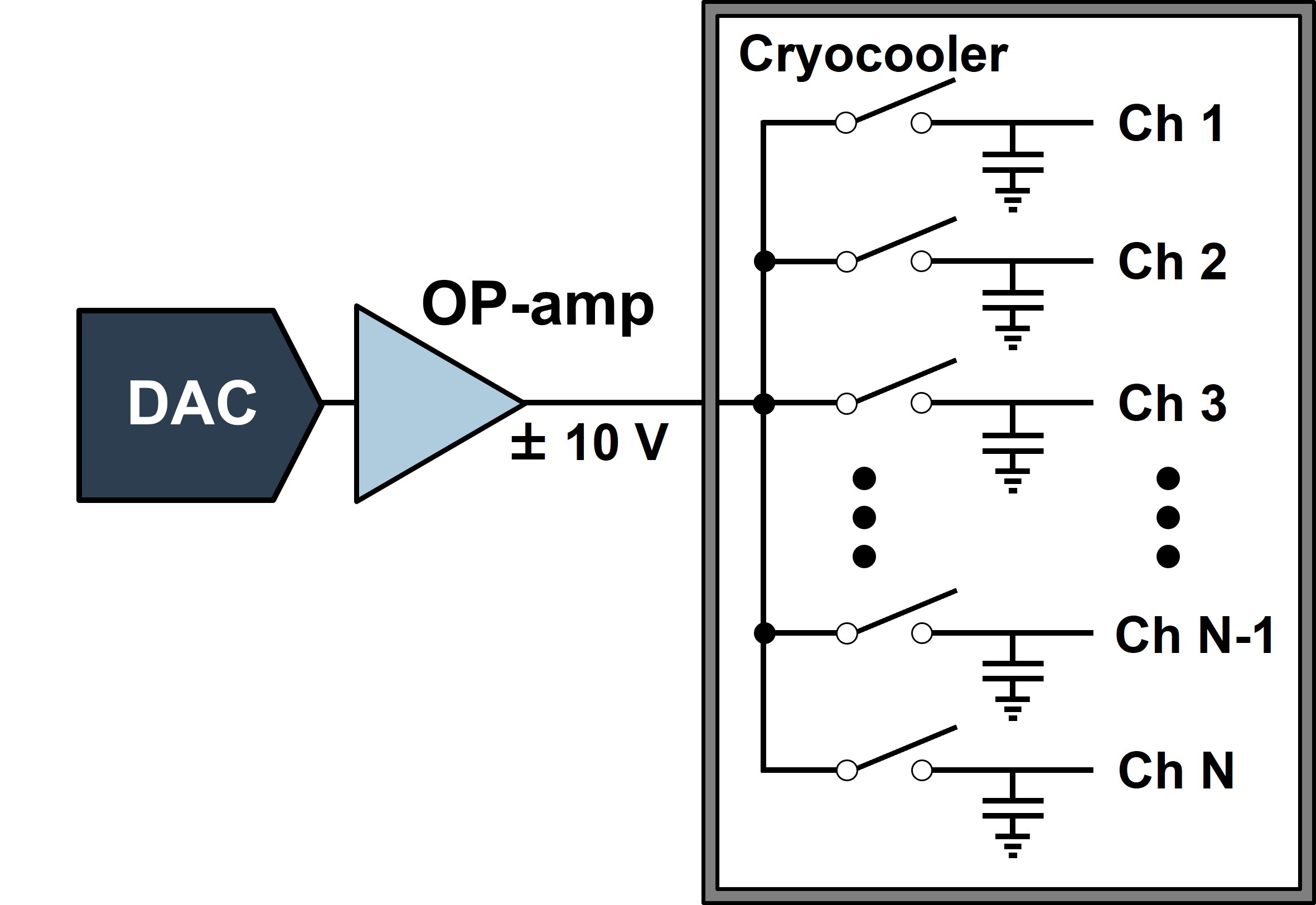}
    \caption{
    Schematic circuit architecture enabling TDM-based voltage control in a cryogenic environment.
    }
    \label{Fig:cryo}
\end{figure}

In summary, we have demonstrated the successful trapping and shuttling of a single $^{40}\rm{Ca}^+$ ion using the proposed TDM-based voltage control scheme.
This method enables the control of ten DC electrodes in a surface trap using only two DACs, thereby eliminating the need for ten independent voltage sources. 
While the current study establishes a functional proof of concept, several challenges must be addressed for future implementation and system scaling.

One primary challenge is the integration of the demultiplexing unit with the vacuum system.
In the present setup, all control electronics are located outside the vacuum chamber. 
This limits scalability due to the finite number of available vacuum feedthroughs.
Although external placement of the electronics was sufficient to demonstrate compatibility with ion-trap operation, in-vacuum integration will be necessary for scaling to larger systems.

Here, we discuss the feasibility of introducing the TDM-based voltage control scheme into a cryogenic environment, taking the 40~K stage as an example.
The dominant source of power dissipation is the final-stage OP-amp.
For the THS4631 used in this work, the quiescent current is 12.5~mA, giving a static dissipation of 375~mW at $\pm15$~V voltage supply.
With a cooling power of $\sim40$~W at the 40~K stage of a commercial Gifford–McMahon cryocooler, this would allow integration of about 100 channels.
This estimate assumes a high-slew-rate amplifier (1000~V/$\mu$s). However, such specifications are not strictly required for ion-trap operation, and substantially lower-slew-rate devices may be considered. 
For example, the LT1351 (Analog Devices, Inc., 200~V/$\mu$s slew rate) exhibits a quiescent current of only 0.25~mA, enabling integration of ~5000 channels.
Further scalability may be achieved with the architecture shown in Fig.~\ref{Fig:cryo}, where signals are pre-amplified at a higher-temperature stage. 
In this configuration, switches must tolerate voltage amplitudes of approximately 10~V and remain operable at cryogenic temperatures. 
Although the attainable switching speed would be limited, the need to place OP-amps directly inside the cryostat is eliminated, thereby reducing the thermal load.
If suitable cryogenic switches operable at 4~K become available, realization of this scheme at the 4~K stage would also be plausible.

Another open issue concerns the potential impact of this control method on motional coherence.
Since the demultiplexing scheme relies on capacitors to store charge, gradual leakage can cause voltage drift.
Fluctuations in the trapping potential may, in turn, affect the phonon coherence of the trapped ions.
While this work does not address this aspect, it should be investigated in future studies.

Finally, although the present study demonstrates the trapping of a single ion, a complete QCCD architecture requires more advanced operations, including linear ion transport, swapping, splitting, and merging~\cite{kaushal2020shuttling, pino2021demonstration, ruster2014experimental, kaufmann2014dynamics, fallek2024rapid, splatt2009deterministic, kaufmann2017fast, van2020coherent}.
These operations are expected to be compatible with the proposed TDM-based voltage control scheme.

\begin{acknowledgments}
This work was supported by MEXT Q-LEAP (Grant Number JPMXS0120319794), JST (Grant Number JPMJPF2014), and JST Moonshot R\&D (Grant Numbers JPMJMS226A, JPMJMS2063).
\end{acknowledgments}

\section*{Author Contributions}
\textbf{Ryutaro~Ohira:} Conceptualization (equal); Data curation (lead); Formal analysis (lead); Methodology (equal); Investigation (equal); Project administration (lead); Supervision (lead); Visualization (lead); Writing/Original Draft Preparation (lead); Writing/Review \& Editing (lead).
\textbf{Masanari~Miyamoto} Investigation (equal); Resources (supporting); Writing/Review \& Editing (supporting).
\textbf{Shinichi~Morisaka} Conceptualization (equal); Methodology (equal); Investigation (equal); Software(equal).
\textbf{Ippei~Nakamura} Conceptualization (equal).
\textbf{Atsushi~Noguchi} Conceptualization (equal); Funding Acquisition (equal); Writing/Review \& Editing (supporting).
\textbf{Utako~Tanaka} Investigation (equal); Resources (equal); Funding Acquisition (equal); Writing/Review \& Editing (supporting).
\textbf{Takefumi~Miyoshi} Conceptualization (equal); Methodology (equal); Investigation (equal); Resources (equal); Funding Acquisition (equal); Software (equal); Writing/Review \& Editing (supporting).

\section*{Data Availability}
The data that support the findings of this study are available from the corresponding author upon reasonable request.

\bibliographystyle{apsrev4-2}
\bibliography{ref}

\begin{thebibliography}{27}%
\makeatletter
\providecommand \@ifxundefined [1]{%
 \@ifx{#1\undefined}
}%
\providecommand \@ifnum [1]{%
 \ifnum #1\expandafter \@firstoftwo
 \else \expandafter \@secondoftwo
 \fi
}%
\providecommand \@ifx [1]{%
 \ifx #1\expandafter \@firstoftwo
 \else \expandafter \@secondoftwo
 \fi
}%
\providecommand \natexlab [1]{#1}%
\providecommand \enquote  [1]{``#1''}%
\providecommand \bibnamefont  [1]{#1}%
\providecommand \bibfnamefont [1]{#1}%
\providecommand \citenamefont [1]{#1}%
\providecommand \href@noop [0]{\@secondoftwo}%
\providecommand \href [0]{\begingroup \@sanitize@url \@href}%
\providecommand \@href[1]{\@@startlink{#1}\@@href}%
\providecommand \@@href[1]{\endgroup#1\@@endlink}%
\providecommand \@sanitize@url [0]{\catcode `\\12\catcode `\$12\catcode `\&12\catcode `\#12\catcode `\^12\catcode `\_12\catcode `\%12\relax}%
\providecommand \@@startlink[1]{}%
\providecommand \@@endlink[0]{}%
\providecommand \url  [0]{\begingroup\@sanitize@url \@url }%
\providecommand \@url [1]{\endgroup\@href {#1}{\urlprefix }}%
\providecommand \urlprefix  [0]{URL }%
\providecommand \Eprint [0]{\href }%
\providecommand \doibase [0]{https://doi.org/}%
\providecommand \selectlanguage [0]{\@gobble}%
\providecommand \bibinfo  [0]{\@secondoftwo}%
\providecommand \bibfield  [0]{\@secondoftwo}%
\providecommand \translation [1]{[#1]}%
\providecommand \BibitemOpen [0]{}%
\providecommand \bibitemStop [0]{}%
\providecommand \bibitemNoStop [0]{.\EOS\space}%
\providecommand \EOS [0]{\spacefactor3000\relax}%
\providecommand \BibitemShut  [1]{\csname bibitem#1\endcsname}%
\let\auto@bib@innerbib\@empty
\bibitem [{\citenamefont {Wineland}\ \emph {et~al.}(1998)\citenamefont {Wineland}, \citenamefont {Monroe}, \citenamefont {Itano}, \citenamefont {Leibfried}, \citenamefont {King},\ and\ \citenamefont {Meekhof}}]{wineland1998experimental}%
  \BibitemOpen
  \bibfield  {author} {\bibinfo {author} {\bibfnamefont {D.~J.}\ \bibnamefont {Wineland}}, \bibinfo {author} {\bibfnamefont {C.}~\bibnamefont {Monroe}}, \bibinfo {author} {\bibfnamefont {W.~M.}\ \bibnamefont {Itano}}, \bibinfo {author} {\bibfnamefont {D.}~\bibnamefont {Leibfried}}, \bibinfo {author} {\bibfnamefont {B.~E.}\ \bibnamefont {King}},\ and\ \bibinfo {author} {\bibfnamefont {D.~M.}\ \bibnamefont {Meekhof}},\ }\href@noop {} {\bibfield  {journal} {\bibinfo  {journal} {J. Res. Natl. Inst. Stand. Technol.}\ }\textbf {\bibinfo {volume} {103}},\ \bibinfo {pages} {259} (\bibinfo {year} {1998})}\BibitemShut {NoStop}%
\bibitem [{\citenamefont {Kielpinski}\ \emph {et~al.}(2002)\citenamefont {Kielpinski}, \citenamefont {Monroe},\ and\ \citenamefont {Wineland}}]{kielpinski2002architecture}%
  \BibitemOpen
  \bibfield  {author} {\bibinfo {author} {\bibfnamefont {D.}~\bibnamefont {Kielpinski}}, \bibinfo {author} {\bibfnamefont {C.}~\bibnamefont {Monroe}},\ and\ \bibinfo {author} {\bibfnamefont {D.~J.}\ \bibnamefont {Wineland}},\ }\href@noop {} {\bibfield  {journal} {\bibinfo  {journal} {Nature}\ }\textbf {\bibinfo {volume} {417}},\ \bibinfo {pages} {709} (\bibinfo {year} {2002})}\BibitemShut {NoStop}%
\bibitem [{\citenamefont {Lekitsch}\ \emph {et~al.}(2017)\citenamefont {Lekitsch}, \citenamefont {Weidt}, \citenamefont {Fowler}, \citenamefont {M{\o}lmer}, \citenamefont {Devitt}, \citenamefont {Wunderlich},\ and\ \citenamefont {Hensinger}}]{lekitsch2017blueprint}%
  \BibitemOpen
  \bibfield  {author} {\bibinfo {author} {\bibfnamefont {B.}~\bibnamefont {Lekitsch}}, \bibinfo {author} {\bibfnamefont {S.}~\bibnamefont {Weidt}}, \bibinfo {author} {\bibfnamefont {A.~G.}\ \bibnamefont {Fowler}}, \bibinfo {author} {\bibfnamefont {K.}~\bibnamefont {M{\o}lmer}}, \bibinfo {author} {\bibfnamefont {S.~J.}\ \bibnamefont {Devitt}}, \bibinfo {author} {\bibfnamefont {C.}~\bibnamefont {Wunderlich}},\ and\ \bibinfo {author} {\bibfnamefont {W.~K.}\ \bibnamefont {Hensinger}},\ }\href@noop {} {\bibfield  {journal} {\bibinfo  {journal} {Sci. Adv.}\ }\textbf {\bibinfo {volume} {3}},\ \bibinfo {pages} {e1601540} (\bibinfo {year} {2017})}\BibitemShut {NoStop}%
\bibitem [{\citenamefont {Kaushal}\ \emph {et~al.}(2020)\citenamefont {Kaushal}, \citenamefont {Lekitsch}, \citenamefont {Stahl}, \citenamefont {Hilder}, \citenamefont {Pijn}, \citenamefont {Schmiegelow}, \citenamefont {Bermudez}, \citenamefont {M{\"u}ller}, \citenamefont {Schmidt-Kaler},\ and\ \citenamefont {Poschinger}}]{kaushal2020shuttling}%
  \BibitemOpen
  \bibfield  {author} {\bibinfo {author} {\bibfnamefont {V.}~\bibnamefont {Kaushal}}, \bibinfo {author} {\bibfnamefont {B.}~\bibnamefont {Lekitsch}}, \bibinfo {author} {\bibfnamefont {A.}~\bibnamefont {Stahl}}, \bibinfo {author} {\bibfnamefont {J.}~\bibnamefont {Hilder}}, \bibinfo {author} {\bibfnamefont {D.}~\bibnamefont {Pijn}}, \bibinfo {author} {\bibfnamefont {C.}~\bibnamefont {Schmiegelow}}, \bibinfo {author} {\bibfnamefont {A.}~\bibnamefont {Bermudez}}, \bibinfo {author} {\bibfnamefont {M.}~\bibnamefont {M{\"u}ller}}, \bibinfo {author} {\bibfnamefont {F.}~\bibnamefont {Schmidt-Kaler}},\ and\ \bibinfo {author} {\bibfnamefont {U.}~\bibnamefont {Poschinger}},\ }\href@noop {} {\bibfield  {journal} {\bibinfo  {journal} {AVS Quantum Sci.}\ }\textbf {\bibinfo {volume} {2}} (\bibinfo {year} {2020})}\BibitemShut {NoStop}%
\bibitem [{\citenamefont {Pino}\ \emph {et~al.}(2021)\citenamefont {Pino}, \citenamefont {Dreiling}, \citenamefont {Figgatt}, \citenamefont {Gaebler}, \citenamefont {Moses}, \citenamefont {Allman}, \citenamefont {Baldwin}, \citenamefont {Foss-Feig}, \citenamefont {Hayes}, \citenamefont {Mayer} \emph {et~al.}}]{pino2021demonstration}%
  \BibitemOpen
  \bibfield  {author} {\bibinfo {author} {\bibfnamefont {J.~M.}\ \bibnamefont {Pino}}, \bibinfo {author} {\bibfnamefont {J.~M.}\ \bibnamefont {Dreiling}}, \bibinfo {author} {\bibfnamefont {C.}~\bibnamefont {Figgatt}}, \bibinfo {author} {\bibfnamefont {J.~P.}\ \bibnamefont {Gaebler}}, \bibinfo {author} {\bibfnamefont {S.~A.}\ \bibnamefont {Moses}}, \bibinfo {author} {\bibfnamefont {M.}~\bibnamefont {Allman}}, \bibinfo {author} {\bibfnamefont {C.}~\bibnamefont {Baldwin}}, \bibinfo {author} {\bibfnamefont {M.}~\bibnamefont {Foss-Feig}}, \bibinfo {author} {\bibfnamefont {D.}~\bibnamefont {Hayes}}, \bibinfo {author} {\bibfnamefont {K.}~\bibnamefont {Mayer}}, \emph {et~al.},\ }\href@noop {} {\bibfield  {journal} {\bibinfo  {journal} {Nature}\ }\textbf {\bibinfo {volume} {592}},\ \bibinfo {pages} {209} (\bibinfo {year} {2021})}\BibitemShut {NoStop}%
\bibitem [{\citenamefont {Moses}\ \emph {et~al.}(2023)\citenamefont {Moses}, \citenamefont {Baldwin}, \citenamefont {Allman}, \citenamefont {Ancona}, \citenamefont {Ascarrunz}, \citenamefont {Barnes}, \citenamefont {Bartolotta}, \citenamefont {Bjork}, \citenamefont {Blanchard}, \citenamefont {Bohn} \emph {et~al.}}]{moses2023race}%
  \BibitemOpen
  \bibfield  {author} {\bibinfo {author} {\bibfnamefont {S.~A.}\ \bibnamefont {Moses}}, \bibinfo {author} {\bibfnamefont {C.~H.}\ \bibnamefont {Baldwin}}, \bibinfo {author} {\bibfnamefont {M.~S.}\ \bibnamefont {Allman}}, \bibinfo {author} {\bibfnamefont {R.}~\bibnamefont {Ancona}}, \bibinfo {author} {\bibfnamefont {L.}~\bibnamefont {Ascarrunz}}, \bibinfo {author} {\bibfnamefont {C.}~\bibnamefont {Barnes}}, \bibinfo {author} {\bibfnamefont {J.}~\bibnamefont {Bartolotta}}, \bibinfo {author} {\bibfnamefont {B.}~\bibnamefont {Bjork}}, \bibinfo {author} {\bibfnamefont {P.}~\bibnamefont {Blanchard}}, \bibinfo {author} {\bibfnamefont {M.}~\bibnamefont {Bohn}}, \emph {et~al.},\ }\href@noop {} {\bibfield  {journal} {\bibinfo  {journal} {Phys. Rev. X}\ }\textbf {\bibinfo {volume} {13}},\ \bibinfo {pages} {041052} (\bibinfo {year} {2023})}\BibitemShut {NoStop}%
\bibitem [{\citenamefont {Mordini}\ \emph {et~al.}(2025)\citenamefont {Mordini}, \citenamefont {Ricci~Vasquez}, \citenamefont {Motohashi}, \citenamefont {M{\"u}ller}, \citenamefont {Malinowski}, \citenamefont {Zhang}, \citenamefont {Mehta}, \citenamefont {Kienzler},\ and\ \citenamefont {Home}}]{mordini2025multizone}%
  \BibitemOpen
  \bibfield  {author} {\bibinfo {author} {\bibfnamefont {C.}~\bibnamefont {Mordini}}, \bibinfo {author} {\bibfnamefont {A.}~\bibnamefont {Ricci~Vasquez}}, \bibinfo {author} {\bibfnamefont {Y.}~\bibnamefont {Motohashi}}, \bibinfo {author} {\bibfnamefont {M.}~\bibnamefont {M{\"u}ller}}, \bibinfo {author} {\bibfnamefont {M.}~\bibnamefont {Malinowski}}, \bibinfo {author} {\bibfnamefont {C.}~\bibnamefont {Zhang}}, \bibinfo {author} {\bibfnamefont {K.~K.}\ \bibnamefont {Mehta}}, \bibinfo {author} {\bibfnamefont {D.}~\bibnamefont {Kienzler}},\ and\ \bibinfo {author} {\bibfnamefont {J.~P.}\ \bibnamefont {Home}},\ }\href@noop {} {\bibfield  {journal} {\bibinfo  {journal} {Phys. Rev. X}\ }\textbf {\bibinfo {volume} {15}},\ \bibinfo {pages} {011040} (\bibinfo {year} {2025})}\BibitemShut {NoStop}%
\bibitem [{\citenamefont {Malinowski}\ \emph {et~al.}(2023)\citenamefont {Malinowski}, \citenamefont {Allcock},\ and\ \citenamefont {Ballance}}]{malinowski2023wire}%
  \BibitemOpen
  \bibfield  {author} {\bibinfo {author} {\bibfnamefont {M.}~\bibnamefont {Malinowski}}, \bibinfo {author} {\bibfnamefont {D.}~\bibnamefont {Allcock}},\ and\ \bibinfo {author} {\bibfnamefont {C.}~\bibnamefont {Ballance}},\ }\href@noop {} {\bibfield  {journal} {\bibinfo  {journal} {PRX Quantum}\ }\textbf {\bibinfo {volume} {4}},\ \bibinfo {pages} {040313} (\bibinfo {year} {2023})}\BibitemShut {NoStop}%
\bibitem [{\citenamefont {Delaney}\ \emph {et~al.}(2024)\citenamefont {Delaney}, \citenamefont {Sletten}, \citenamefont {Cich}, \citenamefont {Estey}, \citenamefont {Fabrikant}, \citenamefont {Hayes}, \citenamefont {Hoffman}, \citenamefont {Hostetter}, \citenamefont {Langer}, \citenamefont {Moses} \emph {et~al.}}]{delaney2024scalable}%
  \BibitemOpen
  \bibfield  {author} {\bibinfo {author} {\bibfnamefont {R.~D.}\ \bibnamefont {Delaney}}, \bibinfo {author} {\bibfnamefont {L.~R.}\ \bibnamefont {Sletten}}, \bibinfo {author} {\bibfnamefont {M.~J.}\ \bibnamefont {Cich}}, \bibinfo {author} {\bibfnamefont {B.}~\bibnamefont {Estey}}, \bibinfo {author} {\bibfnamefont {M.~I.}\ \bibnamefont {Fabrikant}}, \bibinfo {author} {\bibfnamefont {D.}~\bibnamefont {Hayes}}, \bibinfo {author} {\bibfnamefont {I.~M.}\ \bibnamefont {Hoffman}}, \bibinfo {author} {\bibfnamefont {J.}~\bibnamefont {Hostetter}}, \bibinfo {author} {\bibfnamefont {C.}~\bibnamefont {Langer}}, \bibinfo {author} {\bibfnamefont {S.~A.}\ \bibnamefont {Moses}}, \emph {et~al.},\ }\href@noop {} {\bibfield  {journal} {\bibinfo  {journal} {Phys. Rev. X}\ }\textbf {\bibinfo {volume} {14}},\ \bibinfo {pages} {041028} (\bibinfo {year} {2024})}\BibitemShut {NoStop}%
\bibitem [{\citenamefont {Guise}\ \emph {et~al.}(2014)\citenamefont {Guise}, \citenamefont {Fallek}, \citenamefont {Hayden}, \citenamefont {Pai}, \citenamefont {Volin}, \citenamefont {Brown}, \citenamefont {Merrill}, \citenamefont {Harter}, \citenamefont {Amini}, \citenamefont {Lust} \emph {et~al.}}]{guise2014vacuum}%
  \BibitemOpen
  \bibfield  {author} {\bibinfo {author} {\bibfnamefont {N.~D.}\ \bibnamefont {Guise}}, \bibinfo {author} {\bibfnamefont {S.~D.}\ \bibnamefont {Fallek}}, \bibinfo {author} {\bibfnamefont {H.}~\bibnamefont {Hayden}}, \bibinfo {author} {\bibfnamefont {C.}~\bibnamefont {Pai}}, \bibinfo {author} {\bibfnamefont {C.}~\bibnamefont {Volin}}, \bibinfo {author} {\bibfnamefont {K.}~\bibnamefont {Brown}}, \bibinfo {author} {\bibfnamefont {J.}~\bibnamefont {Merrill}}, \bibinfo {author} {\bibfnamefont {A.~W.}\ \bibnamefont {Harter}}, \bibinfo {author} {\bibfnamefont {J.~M.}\ \bibnamefont {Amini}}, \bibinfo {author} {\bibfnamefont {L.~M.}\ \bibnamefont {Lust}}, \emph {et~al.},\ }\href@noop {} {\bibfield  {journal} {\bibinfo  {journal} {Rev. Sci. Instrum.}\ }\textbf {\bibinfo {volume} {85}} (\bibinfo {year} {2014})}\BibitemShut {NoStop}%
\bibitem [{\citenamefont {Stuart}\ \emph {et~al.}(2019)\citenamefont {Stuart}, \citenamefont {Panock}, \citenamefont {Bruzewicz}, \citenamefont {Sedlacek}, \citenamefont {McConnell}, \citenamefont {Chuang}, \citenamefont {Sage},\ and\ \citenamefont {Chiaverini}}]{stuart2019chip}%
  \BibitemOpen
  \bibfield  {author} {\bibinfo {author} {\bibfnamefont {J.}~\bibnamefont {Stuart}}, \bibinfo {author} {\bibfnamefont {R.}~\bibnamefont {Panock}}, \bibinfo {author} {\bibfnamefont {C.}~\bibnamefont {Bruzewicz}}, \bibinfo {author} {\bibfnamefont {J.}~\bibnamefont {Sedlacek}}, \bibinfo {author} {\bibfnamefont {R.}~\bibnamefont {McConnell}}, \bibinfo {author} {\bibfnamefont {I.}~\bibnamefont {Chuang}}, \bibinfo {author} {\bibfnamefont {J.}~\bibnamefont {Sage}},\ and\ \bibinfo {author} {\bibfnamefont {J.}~\bibnamefont {Chiaverini}},\ }\href@noop {} {\bibfield  {journal} {\bibinfo  {journal} {Phys. Rev. Appl.}\ }\textbf {\bibinfo {volume} {11}},\ \bibinfo {pages} {024010} (\bibinfo {year} {2019})}\BibitemShut {NoStop}%
\bibitem [{\citenamefont {Ohira}\ \emph {et~al.}(2025)\citenamefont {Ohira}, \citenamefont {Morisaka}, \citenamefont {Nakamura}, \citenamefont {Noguchi},\ and\ \citenamefont {Miyoshi}}]{ohira2025multiplexed}%
  \BibitemOpen
  \bibfield  {author} {\bibinfo {author} {\bibfnamefont {R.}~\bibnamefont {Ohira}}, \bibinfo {author} {\bibfnamefont {S.}~\bibnamefont {Morisaka}}, \bibinfo {author} {\bibfnamefont {I.}~\bibnamefont {Nakamura}}, \bibinfo {author} {\bibfnamefont {A.}~\bibnamefont {Noguchi}},\ and\ \bibinfo {author} {\bibfnamefont {T.}~\bibnamefont {Miyoshi}},\ }\href@noop {} {\bibfield  {journal} {\bibinfo  {journal} {arXiv:2504.01815}\ } (\bibinfo {year} {2025})}\BibitemShut {NoStop}%
\bibitem [{\citenamefont {Blakestad}(2010)}]{blakestad2010transport}%
  \BibitemOpen
  \bibfield  {author} {\bibinfo {author} {\bibfnamefont {R.~B.}\ \bibnamefont {Blakestad}},\ }\emph {\bibinfo {title} {Transport of trapped-ion qubits within a scalable quantum processor}},\ \href@noop {} {Ph.D. thesis},\ \bibinfo  {school} {University of Colorado at Boulder} (\bibinfo {year} {2010})\BibitemShut {NoStop}%
\bibitem [{\citenamefont {Blakestad}\ \emph {et~al.}(2011)\citenamefont {Blakestad}, \citenamefont {Ospelkaus}, \citenamefont {VanDevender}, \citenamefont {Wesenberg}, \citenamefont {Biercuk}, \citenamefont {Leibfried},\ and\ \citenamefont {Wineland}}]{blakestad2011near}%
  \BibitemOpen
  \bibfield  {author} {\bibinfo {author} {\bibfnamefont {R.}~\bibnamefont {Blakestad}}, \bibinfo {author} {\bibfnamefont {C.}~\bibnamefont {Ospelkaus}}, \bibinfo {author} {\bibfnamefont {A.}~\bibnamefont {VanDevender}}, \bibinfo {author} {\bibfnamefont {J.}~\bibnamefont {Wesenberg}}, \bibinfo {author} {\bibfnamefont {M.}~\bibnamefont {Biercuk}}, \bibinfo {author} {\bibfnamefont {D.}~\bibnamefont {Leibfried}},\ and\ \bibinfo {author} {\bibfnamefont {D.~J.}\ \bibnamefont {Wineland}},\ }\href@noop {} {\bibfield  {journal} {\bibinfo  {journal} {Phys. Rev. A}\ }\textbf {\bibinfo {volume} {84}},\ \bibinfo {pages} {032314} (\bibinfo {year} {2011})}\BibitemShut {NoStop}%
\bibitem [{\citenamefont {Akhtar}\ \emph {et~al.}(2023)\citenamefont {Akhtar}, \citenamefont {Bonus}, \citenamefont {Lebrun-Gallagher}, \citenamefont {Johnson}, \citenamefont {Siegele-Brown}, \citenamefont {Hong}, \citenamefont {Hile}, \citenamefont {Kulmiya}, \citenamefont {Weidt},\ and\ \citenamefont {Hensinger}}]{akhtar2023high}%
  \BibitemOpen
  \bibfield  {author} {\bibinfo {author} {\bibfnamefont {M.}~\bibnamefont {Akhtar}}, \bibinfo {author} {\bibfnamefont {F.}~\bibnamefont {Bonus}}, \bibinfo {author} {\bibfnamefont {F.}~\bibnamefont {Lebrun-Gallagher}}, \bibinfo {author} {\bibfnamefont {N.}~\bibnamefont {Johnson}}, \bibinfo {author} {\bibfnamefont {M.}~\bibnamefont {Siegele-Brown}}, \bibinfo {author} {\bibfnamefont {S.}~\bibnamefont {Hong}}, \bibinfo {author} {\bibfnamefont {S.}~\bibnamefont {Hile}}, \bibinfo {author} {\bibfnamefont {S.}~\bibnamefont {Kulmiya}}, \bibinfo {author} {\bibfnamefont {S.}~\bibnamefont {Weidt}},\ and\ \bibinfo {author} {\bibfnamefont {W.}~\bibnamefont {Hensinger}},\ }\href@noop {} {\bibfield  {journal} {\bibinfo  {journal} {Nat. Commun.}\ }\textbf {\bibinfo {volume} {14}},\ \bibinfo {pages} {531} (\bibinfo {year} {2023})}\BibitemShut {NoStop}%
\bibitem [{\citenamefont {Miyoshi}\ \emph {et~al.}(2025)\citenamefont {Miyoshi}, \citenamefont {Koike}, \citenamefont {Morisaka}, \citenamefont {Sugita}, \citenamefont {Sumida}, \citenamefont {Tabuchi}, \citenamefont {Negoro}, \citenamefont {Shiomi}, \citenamefont {Nakamura}, \citenamefont {Tomita} \emph {et~al.}}]{miyoshi2025toward}%
  \BibitemOpen
  \bibfield  {author} {\bibinfo {author} {\bibfnamefont {T.}~\bibnamefont {Miyoshi}}, \bibinfo {author} {\bibfnamefont {K.}~\bibnamefont {Koike}}, \bibinfo {author} {\bibfnamefont {S.}~\bibnamefont {Morisaka}}, \bibinfo {author} {\bibfnamefont {Y.}~\bibnamefont {Sugita}}, \bibinfo {author} {\bibfnamefont {T.}~\bibnamefont {Sumida}}, \bibinfo {author} {\bibfnamefont {Y.}~\bibnamefont {Tabuchi}}, \bibinfo {author} {\bibfnamefont {M.}~\bibnamefont {Negoro}}, \bibinfo {author} {\bibfnamefont {H.}~\bibnamefont {Shiomi}}, \bibinfo {author} {\bibfnamefont {I.}~\bibnamefont {Nakamura}}, \bibinfo {author} {\bibfnamefont {T.}~\bibnamefont {Tomita}}, \emph {et~al.},\ }in\ \href@noop {} {\emph {\bibinfo {booktitle} {2025 IEEE International Conference on Consumer Electronics (ICCE)}}}\ (\bibinfo {organization} {IEEE},\ \bibinfo {year} {2025})\ pp.\ \bibinfo {pages} {1--5}\BibitemShut {NoStop}%
\bibitem [{Note1()}]{Note1}%
  \BibitemOpen
  \bibinfo {note} {Zynq UltraScale+ MPSoC ZCU106 Evaluation Kit, Advanced Micro Devices, Inc.}\BibitemShut {Stop}%
\bibitem [{\citenamefont {{Texas Instruments}}()}]{SN74AUC1G66-datasheet}%
  \BibitemOpen
  \bibfield  {author} {\bibinfo {author} {\bibnamefont {{Texas Instruments}}},\ }\href@noop {} {\bibinfo {title} {{SN74AUC1G66 datasheet (Rev. L)}}},\ \bibinfo {note} {accessed: 2025-07-22}\BibitemShut {NoStop}%
\bibitem [{Note2()}]{Note2}%
  \BibitemOpen
  \bibinfo {note} {$R_0$ and $R_1$ are general-purpose chip resistors, RK73B1JTTD822J and RK73B1JTTD102J, respectively, manufactured by KOA Speer Electronics, Inc. According to their datasheet, both resistors have a tolerance of $\pm 5\%$.}\BibitemShut {Stop}%
\bibitem [{\citenamefont {Miyamoto}\ \emph {et~al.}(2025)\citenamefont {Miyamoto}, \citenamefont {Higuchi}, \citenamefont {Furusawa}, \citenamefont {Sekine}, \citenamefont {Hayasaka},\ and\ \citenamefont {Tanaka}}]{miyamoto2025isotope}%
  \BibitemOpen
  \bibfield  {author} {\bibinfo {author} {\bibfnamefont {M.}~\bibnamefont {Miyamoto}}, \bibinfo {author} {\bibfnamefont {T.}~\bibnamefont {Higuchi}}, \bibinfo {author} {\bibfnamefont {K.}~\bibnamefont {Furusawa}}, \bibinfo {author} {\bibfnamefont {N.}~\bibnamefont {Sekine}}, \bibinfo {author} {\bibfnamefont {K.}~\bibnamefont {Hayasaka}},\ and\ \bibinfo {author} {\bibfnamefont {U.}~\bibnamefont {Tanaka}},\ }\href@noop {} {\bibfield  {journal} {\bibinfo  {journal} {arXiv:2505.05002}\ } (\bibinfo {year} {2025})}\BibitemShut {NoStop}%
\bibitem [{\citenamefont {Oshio}\ \emph {et~al.}(2025)\citenamefont {Oshio}, \citenamefont {Nishimoto}, \citenamefont {Higuchi}, \citenamefont {Hayasaka}, \citenamefont {Koike}, \citenamefont {Morisaka}, \citenamefont {Miyoshi}, \citenamefont {Ohira},\ and\ \citenamefont {Tanaka}}]{oshio2025development}%
  \BibitemOpen
  \bibfield  {author} {\bibinfo {author} {\bibfnamefont {T.}~\bibnamefont {Oshio}}, \bibinfo {author} {\bibfnamefont {R.}~\bibnamefont {Nishimoto}}, \bibinfo {author} {\bibfnamefont {T.}~\bibnamefont {Higuchi}}, \bibinfo {author} {\bibfnamefont {K.}~\bibnamefont {Hayasaka}}, \bibinfo {author} {\bibfnamefont {K.}~\bibnamefont {Koike}}, \bibinfo {author} {\bibfnamefont {S.}~\bibnamefont {Morisaka}}, \bibinfo {author} {\bibfnamefont {T.}~\bibnamefont {Miyoshi}}, \bibinfo {author} {\bibfnamefont {R.}~\bibnamefont {Ohira}},\ and\ \bibinfo {author} {\bibfnamefont {U.}~\bibnamefont {Tanaka}},\ }\href@noop {} {\bibfield  {journal} {\bibinfo  {journal} {J. Appl. Phys.}\ }\textbf {\bibinfo {volume} {137}},\ \bibinfo {pages} {144401} (\bibinfo {year} {2025})}\BibitemShut {NoStop}%
\bibitem [{\citenamefont {Ruster}\ \emph {et~al.}(2014)\citenamefont {Ruster}, \citenamefont {Warschburger}, \citenamefont {Kaufmann}, \citenamefont {Schmiegelow}, \citenamefont {Walther}, \citenamefont {Hettrich}, \citenamefont {Pfister}, \citenamefont {Kaushal}, \citenamefont {Schmidt-Kaler},\ and\ \citenamefont {Poschinger}}]{ruster2014experimental}%
  \BibitemOpen
  \bibfield  {author} {\bibinfo {author} {\bibfnamefont {T.}~\bibnamefont {Ruster}}, \bibinfo {author} {\bibfnamefont {C.}~\bibnamefont {Warschburger}}, \bibinfo {author} {\bibfnamefont {H.}~\bibnamefont {Kaufmann}}, \bibinfo {author} {\bibfnamefont {C.~T.}\ \bibnamefont {Schmiegelow}}, \bibinfo {author} {\bibfnamefont {A.}~\bibnamefont {Walther}}, \bibinfo {author} {\bibfnamefont {M.}~\bibnamefont {Hettrich}}, \bibinfo {author} {\bibfnamefont {A.}~\bibnamefont {Pfister}}, \bibinfo {author} {\bibfnamefont {V.}~\bibnamefont {Kaushal}}, \bibinfo {author} {\bibfnamefont {F.}~\bibnamefont {Schmidt-Kaler}},\ and\ \bibinfo {author} {\bibfnamefont {U.~G.}\ \bibnamefont {Poschinger}},\ }\href@noop {} {\bibfield  {journal} {\bibinfo  {journal} {Phys. Rev. A}\ }\textbf {\bibinfo {volume} {90}},\ \bibinfo {pages} {033410} (\bibinfo {year} {2014})}\BibitemShut {NoStop}%
\bibitem [{\citenamefont {Kaufmann}\ \emph {et~al.}(2014)\citenamefont {Kaufmann}, \citenamefont {Ruster}, \citenamefont {Schmiegelow}, \citenamefont {Schmidt-Kaler},\ and\ \citenamefont {Poschinger}}]{kaufmann2014dynamics}%
  \BibitemOpen
  \bibfield  {author} {\bibinfo {author} {\bibfnamefont {H.}~\bibnamefont {Kaufmann}}, \bibinfo {author} {\bibfnamefont {T.}~\bibnamefont {Ruster}}, \bibinfo {author} {\bibfnamefont {C.}~\bibnamefont {Schmiegelow}}, \bibinfo {author} {\bibfnamefont {F.}~\bibnamefont {Schmidt-Kaler}},\ and\ \bibinfo {author} {\bibfnamefont {U.}~\bibnamefont {Poschinger}},\ }\href@noop {} {\bibfield  {journal} {\bibinfo  {journal} {New J. Phys.}\ }\textbf {\bibinfo {volume} {16}},\ \bibinfo {pages} {073012} (\bibinfo {year} {2014})}\BibitemShut {NoStop}%
\bibitem [{\citenamefont {Fallek}\ \emph {et~al.}(2024)\citenamefont {Fallek}, \citenamefont {Sandhu}, \citenamefont {McGill}, \citenamefont {Gray}, \citenamefont {Tinkey}, \citenamefont {Clark},\ and\ \citenamefont {Brown}}]{fallek2024rapid}%
  \BibitemOpen
  \bibfield  {author} {\bibinfo {author} {\bibfnamefont {S.~D.}\ \bibnamefont {Fallek}}, \bibinfo {author} {\bibfnamefont {V.~S.}\ \bibnamefont {Sandhu}}, \bibinfo {author} {\bibfnamefont {R.~A.}\ \bibnamefont {McGill}}, \bibinfo {author} {\bibfnamefont {J.~M.}\ \bibnamefont {Gray}}, \bibinfo {author} {\bibfnamefont {H.~N.}\ \bibnamefont {Tinkey}}, \bibinfo {author} {\bibfnamefont {C.~R.}\ \bibnamefont {Clark}},\ and\ \bibinfo {author} {\bibfnamefont {K.~R.}\ \bibnamefont {Brown}},\ }\href@noop {} {\bibfield  {journal} {\bibinfo  {journal} {Nat. Commun.}\ }\textbf {\bibinfo {volume} {15}},\ \bibinfo {pages} {1089} (\bibinfo {year} {2024})}\BibitemShut {NoStop}%
\bibitem [{\citenamefont {Splatt}\ \emph {et~al.}(2009)\citenamefont {Splatt}, \citenamefont {Harlander}, \citenamefont {Brownnutt}, \citenamefont {Z{\"a}hringer}, \citenamefont {Blatt},\ and\ \citenamefont {H{\"a}nsel}}]{splatt2009deterministic}%
  \BibitemOpen
  \bibfield  {author} {\bibinfo {author} {\bibfnamefont {F.}~\bibnamefont {Splatt}}, \bibinfo {author} {\bibfnamefont {M.}~\bibnamefont {Harlander}}, \bibinfo {author} {\bibfnamefont {M.}~\bibnamefont {Brownnutt}}, \bibinfo {author} {\bibfnamefont {F.}~\bibnamefont {Z{\"a}hringer}}, \bibinfo {author} {\bibfnamefont {R.}~\bibnamefont {Blatt}},\ and\ \bibinfo {author} {\bibfnamefont {W.}~\bibnamefont {H{\"a}nsel}},\ }\href@noop {} {\bibfield  {journal} {\bibinfo  {journal} {New J. Phys.}\ }\textbf {\bibinfo {volume} {11}},\ \bibinfo {pages} {103008} (\bibinfo {year} {2009})}\BibitemShut {NoStop}%
\bibitem [{\citenamefont {Kaufmann}\ \emph {et~al.}(2017)\citenamefont {Kaufmann}, \citenamefont {Ruster}, \citenamefont {Schmiegelow}, \citenamefont {Luda}, \citenamefont {Kaushal}, \citenamefont {Schulz}, \citenamefont {von Lindenfels}, \citenamefont {Schmidt-Kaler},\ and\ \citenamefont {Poschinger}}]{kaufmann2017fast}%
  \BibitemOpen
  \bibfield  {author} {\bibinfo {author} {\bibfnamefont {H.}~\bibnamefont {Kaufmann}}, \bibinfo {author} {\bibfnamefont {T.}~\bibnamefont {Ruster}}, \bibinfo {author} {\bibfnamefont {C.~T.}\ \bibnamefont {Schmiegelow}}, \bibinfo {author} {\bibfnamefont {M.~A.}\ \bibnamefont {Luda}}, \bibinfo {author} {\bibfnamefont {V.}~\bibnamefont {Kaushal}}, \bibinfo {author} {\bibfnamefont {J.}~\bibnamefont {Schulz}}, \bibinfo {author} {\bibfnamefont {D.}~\bibnamefont {von Lindenfels}}, \bibinfo {author} {\bibfnamefont {F.}~\bibnamefont {Schmidt-Kaler}},\ and\ \bibinfo {author} {\bibfnamefont {U.~G.}\ \bibnamefont {Poschinger}},\ }\href@noop {} {\bibfield  {journal} {\bibinfo  {journal} {Phys. Rev. A}\ }\textbf {\bibinfo {volume} {95}},\ \bibinfo {pages} {052319} (\bibinfo {year} {2017})}\BibitemShut {NoStop}%
\bibitem [{\citenamefont {van Mourik}\ \emph {et~al.}(2020)\citenamefont {van Mourik}, \citenamefont {Martinez}, \citenamefont {Gerster}, \citenamefont {Hrmo}, \citenamefont {Monz}, \citenamefont {Schindler},\ and\ \citenamefont {Blatt}}]{van2020coherent}%
  \BibitemOpen
  \bibfield  {author} {\bibinfo {author} {\bibfnamefont {M.~W.}\ \bibnamefont {van Mourik}}, \bibinfo {author} {\bibfnamefont {E.~A.}\ \bibnamefont {Martinez}}, \bibinfo {author} {\bibfnamefont {L.}~\bibnamefont {Gerster}}, \bibinfo {author} {\bibfnamefont {P.}~\bibnamefont {Hrmo}}, \bibinfo {author} {\bibfnamefont {T.}~\bibnamefont {Monz}}, \bibinfo {author} {\bibfnamefont {P.}~\bibnamefont {Schindler}},\ and\ \bibinfo {author} {\bibfnamefont {R.}~\bibnamefont {Blatt}},\ }\href@noop {} {\bibfield  {journal} {\bibinfo  {journal} {Phys. Rev. A}\ }\textbf {\bibinfo {volume} {102}},\ \bibinfo {pages} {022611} (\bibinfo {year} {2020})}\BibitemShut {NoStop}%
\end{thebibliography}%

\end{document}